\documentclass[final]{svjour2}
\usepackage{graphicx}
\usepackage{rotating}
\usepackage{amssymb}
\usepackage{mathptmx}
\usepackage[numbers]{natbib}
\makeatletter
\journalname{Journal of Low Temperature Physics}

\bibpunct{}{}{,}{s}{}{,}

\begin{document}

\newcommand{\hdblarrow}{H\makebox[0.9ex][l]{$\downdownarrows$}-}

\title{Development of Multilayer Readout Wiring TES Calorimeter for Future X-ray Missions} 

\author{
S.~Yamada$^{1}$ \and
Y.~Ezoe$^{2}$ \and 
Y. Ishisaki$^{2}$ \and
T.~Ohashi$^{2}$ \and 
N.~Iijima$^{2}$ \and 
K.~Mitsuda$^{3}$ \and
K.~Nagayoshi$^{3}$ \and
H.~Akamatsu$^{4}$ \and 
T.~Morooka$^{5}$ \and
K.~Tanaka$^{6}$
}

\institute{
1.~Cosmic Radiation Laboratory, Institute of Physical and Chemical Research (RIKEN) \\
2.~Tokyo Metropolitan University,~~3.~ISAS/JAXA, ~~ 4.~SRON, \\ 
5.~Seiko Instruments Inc, ~~ 6.~Hitachi High-Tech Science Corporation \\ 
\email{yamada@crab.riken.jp}}

\date{15.7.2013}

\maketitle

\begin{abstract}

We have fabricated multilayer readout wiring Transition Edge Sensors (TES), 
which enable us to realize both large effective area and high-energy
resolution for future X-ray astrophysical missions, such as DIOS. 
By sandwiching a SiO$_2$ insulation layer between Al superconducting signal and return lines, 
self/mutual inductances and self fielding of bias leads are expected to be reduced.
We fabricated 4$\times$4 and 20$\times$20 TES array on the multilayer wiring and tested their performance.
Under the low temperature condition, 
several pixels in the TES array showed sharp superconducting transitions at around $\sim$300 mK. 
We also succeeded in detecting X-ray signals from the 4$\times$4 TES, contrary to the previous results of 20$\times$20 TES. 
We further investigated the reasons for the differences between the 4$\times$4 TES and the 20$\times$20 TES, 
and present future plans for improving the multilayer TES array fabrication. 

\keywords{TES, X-ray astronomy, multilayer wiring}

\end{abstract}

\section{Introduction}


Microcalorimeters, which use small temperature rise ($\sim$ 0.001 K) 
to measure deposited energy, are now becoming the most promising 
and standard detector for the next-generation X-ray astronomical mission, 
such as Micro-X (Figueroa-Feliciano {\it et al.}~2008\cite{Fig08}), 
ASTRO-H SXS (Takahashi {\it et al.}~2010\cite{Takahashi2010}) 
and AXSIO (Bookbinder et al.~2012\cite{Bookbinder2012}).  
Among several types of microcalorimeters, 
the most-established in the X-ray band is 
the Transition-Edge Sensor (TES) thermometer which 
consists of thin-films electrically biased in the superconducting-to-normal transition 
where electrical resistance significantly depends on temperature\cite{KG05}\cite{Smith12}. 

The Japanese future small satellite mission carrying TES calorimeters 
is Diffuse Intergalactic Oxygen Surveyor 
(DIOS; Ohashi {\it et al.} 2010\cite{Ohashi10}), 
aiming for detecting Warm-Hot Intergalactic Medium\cite{Yoshikawa01}. 
In space application of TES used together with X-ray mirrors, 
maximizing X-ray absorber area per given detector area on a focal plane is crucial. 
To realize mission requirement for the wide field of view, 
$S\Omega \sim$ 150 cm$^2$ deg$^2$, 
we have been developing multi-layer wiring TES calorimeter (Ezoe~{\it et al.} 2009\cite{Ezoe09} and 2012\cite{Ezoe12},
Oishi~{\it et al.} 2012\cite{Oishi12}),  based on our standard coplanar-type TES\cite{Aka09}\cite{Ezoe09}. 
 
%

 
One of the multi-layer wiring TES Array (TESA) consists of hot (top layer) and return (bottom layer) wires passing 
parallel to the insulated layer of SiO$_2$, so that wires can be more tightly packed than coplanar wiring, 
also making opening angle wider, as well as reducing cross talks between wires; 
cf., Chervenak~{\it et al.} 2012\cite{Che12}. 
We designed the multi-layer wiring 20$\times$20 pixel TES and studied its basic properties\cite{Ezoe12}, 
but could not obtain an anticipated performance. 
We therefore assessed the performance of 4$\times$4 pixel TESA with multi-layer wiring, 
and compared the results between the 4$\times$4 TESA and the 20$\times$20 TESA. 

\begin{figure}
\begin{center}
\includegraphics[%
  width=0.99\linewidth,
  keepaspectratio]{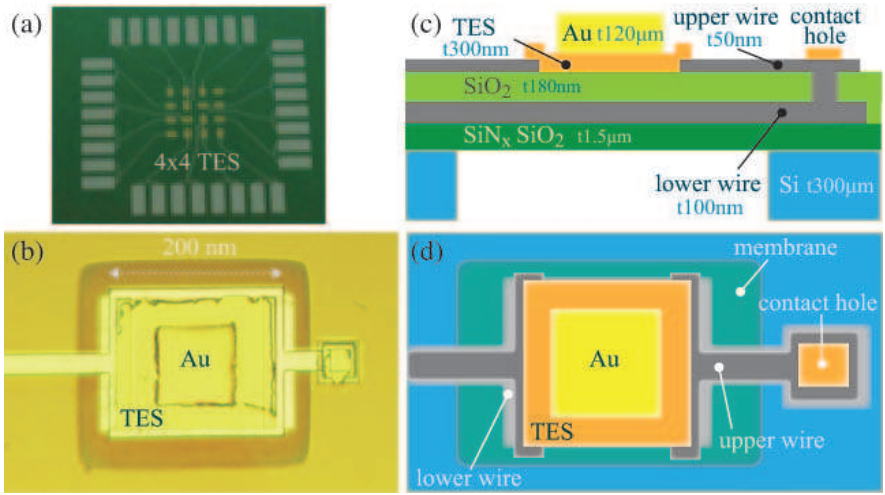}
\end{center}
\caption{(a) Overall picture of the multi-layer wiring 4$\times$4 TESA. (b) Microscope photograph of one TES pixel from the top. (c) Illustrated cross-sectional view of the multi-layer wiring TES. (d) Illustrated top view of the multi-layer wiring TES. Lower wires are behind the upper wires, connecting through the contact hole. (color figure online)}
\label{compare}
\end{figure}

\section{Design of Multi-layer Readout Wiring TES} 

We fabricated a 4$\times$4 TESA on top of the multi-layer wiring. 
Figure 1 shows an overall picture and a closed-up one of the TES. 
Upper (bias) and lower (return) wiring made of Al sandwiches an insulation SiO$_2$ film,
and are linked through a contact hole\cite{Ezoe12}. 
Ti and Au TES layers are 200~$\mu$m on a side by~100 nm and 200~nm thick, respectively. 
Au Absorber is 120~$\mu$m on a side by 1.7~$\mu$m thick;   
upper and lower wire widths are 20~$\mu$m and 30~$\mu$m; 
and heights are 50~nm and 100~nm, respectively. 
Process yield in terms of electrical resistance at room temperature 
were estimated to be $\sim$ 95\%\cite{Oishi12}, 
which were measured by tossing the tips of the manual prober onto each boding pad at room temperature.  


\section{Experimental Results}
\subsection{Setup and Basic properties}
We used the dilution refrigerator (OXFORD Kelvinox25) in Tokyo Metropolitan University, 
which has $\sim$ 25 $\mu$W cooling power with a lowest-achievable temperature of $\sim$ 50~mK. 
The 4$\times$4 TES shown in figure 1 is placed on the detector stage 
with a RuO$_2$ thermometer which is read out and controlled by Picowatt AVS47/TS-530. 
On top of the TES, $^{55}$Fe isotopes is mounted as X-ray sources. 
One of the TES pixel with Au absorber out of 8 pixel is 
voltage-biased and lined to SQUID readout. 

\begin{figure}
\begin{center}
\includegraphics[%
  width=0.99\linewidth,
  keepaspectratio]{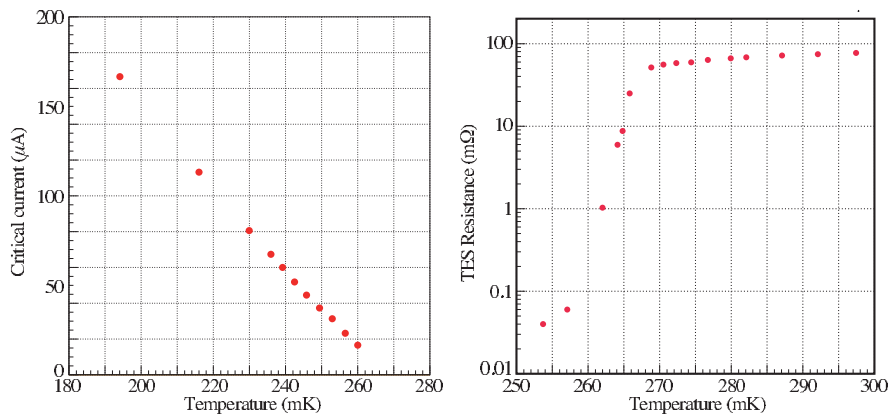}
\end{center}
\caption{(left) Temperature vs. critical current obtained from one of the multi-layer 4$\times$4 TES. (right) Temperature vs. TES resistance taken from the same pixel. (color figure online)}
\label{avsT}
\end{figure}

We measured the critical current ($I_c$) and TES resistance as a function of 
temperature at the detector stage, as shown in figure 2. 
The transition temperature ($T_c$) was $\sim$~265 mK, 
which is 
expected from this TES thicknesses; 
for example, Ti and Au thickness of our coplanar-type TES were $\sim$ 40 nm and $\sim$ 80 nm, respectively.  
We could expect to obtain lower $T_c$ than $\sim$~265 mK when the TES thicknesses are optimized. 

As the temperature is reduced, $I_c$ increases,  
expected from the GL theory. 
The TES resistance ranges from 0.28 m$\Omega$ (super),  which is consistent with zero within error of the measurement,  through 153 m$\Omega$ (normal). 
The obtained residual resistance became lower than $\sim$20 m$\Omega$ obtained from the 20$\times$20 pixel TES\cite{Ezoe12}, which was presumably due to remaining materials on the TES. 
Our coplanar TES measured by Ishisaki {\it et al.} 2007\cite{Ishisaki07} 
showed that $I_c$ was larger than 100 $\mu$A at $\sim$ 90\% normal resistance, 
suggesting that $I_c$ of the multi-wiring TESA is increased by reviewing cleaning processes. 

\subsection{X-ray Irradiation}

We proceeded to measure the X-ray spectrum of the multi-layer wiring TES. 
The thermal bath temperature is set at 185.5~mK with a bias voltage of 2.4~V, 
in order to operate the TES under electro-thermal feedback working. 
We succeeded in obtaining X-ray signal and 
the pulse profile averaged over all the X-ray signals during 
the entire measurement ($\sim$ 10 h) is shown in figure~3 (left). 
The pulse profile of the coplanar wiring TES\cite{Aka09} 
is overlaid, which is taken under the same configuration. 
Apparently, the height of the pulse differs by an order of magnitude. 
The time constants in the multi-layered wiring TES seem 
to be a sum of fast ($\sim$ 0.02 ms) and slow ($\sim$ 0.2 ms) falling components, 
while the fast one is less obvious in the coplanar TES.  
We evaluated the baseline fluctuation by using off-source data, 
and found it to be $\sim$ 40 eV. 

We created optimal filter by collecting X-ray event data and off-source data, 
and then corrected the raw pulse height data for temperature-dependent fluctuation 
by aligning the Mn K$_{\alpha 1}$ peak. 
The obtained spectrum is presented in figure~3 (right).  
The energy resolution is $\sim$ 100 eV at 5.9 keV, 
which would be probably due to its quite low pulse height, 
or might be related to thick TES layers causing the high transition temperature.  

\begin{figure}
\begin{center}
\includegraphics[%
  width=0.99\linewidth,
  keepaspectratio]{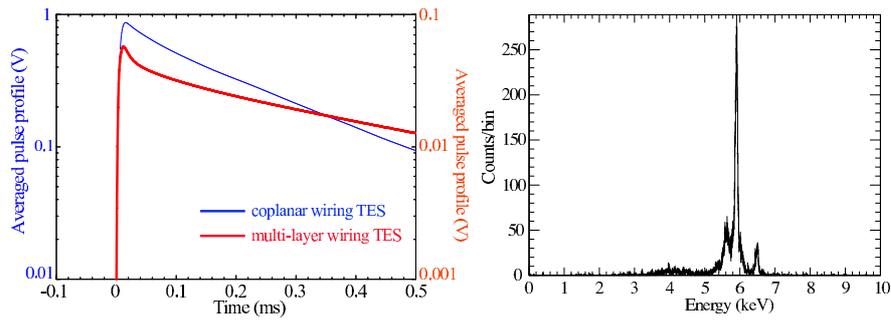}
\end{center}
\caption{(left) Temporal pulse profile of X-ray events taken from one of the multi-layer TESA 
is shown in a solid thick red line. 
As a reference, the pulse profile of the standard coplanar TES 
is overlaid in solid thin blue line. 
(right) X-ray spectrum of of the multi-layer TES pixel obtained by irradiating $^{55}$Fe isotopes.  (color figure online)}
\label{quad}
\end{figure}

\section{Discussion and Conclusion}

We compared 4$\times$4 TESA obtained in this proceeding with 20$\times$20 TESA properties\cite{Ezoe12}. 
The obtained results are summarized in table~1. 
\begin{table}[htbp]
 \caption{Summary of 4$\times$4 TES and 20$\times$20 TES properties.}
 \label{xismode}
 \begin{center}
  \begin{tabular}{ccccccc}
   \hline
  & yield rate & $T_c$ & $I_c$@0.9$R_c$ & $R$ (super) & $R$ (normal) & $\Delta E$ \\ 
\hline	
4$\times$4 TES & $\sim$94\%      & 260 mK &  $\sim$100$\mu$A &  0.28 m$\Omega$ & 153 m$\Omega$ & $\sim$ 100 eV \\
20$\times$20 TES &  $\sim$60\% & 245 mK & $\sim$4$\mu$A &      20 m$\Omega$ & 9 $\Omega$ & -   \\
\hline
\end{tabular}
\end{center}
\end{table}
\begin{figure}
\begin{center}
\includegraphics[%
  width=0.95\linewidth,
  keepaspectratio]{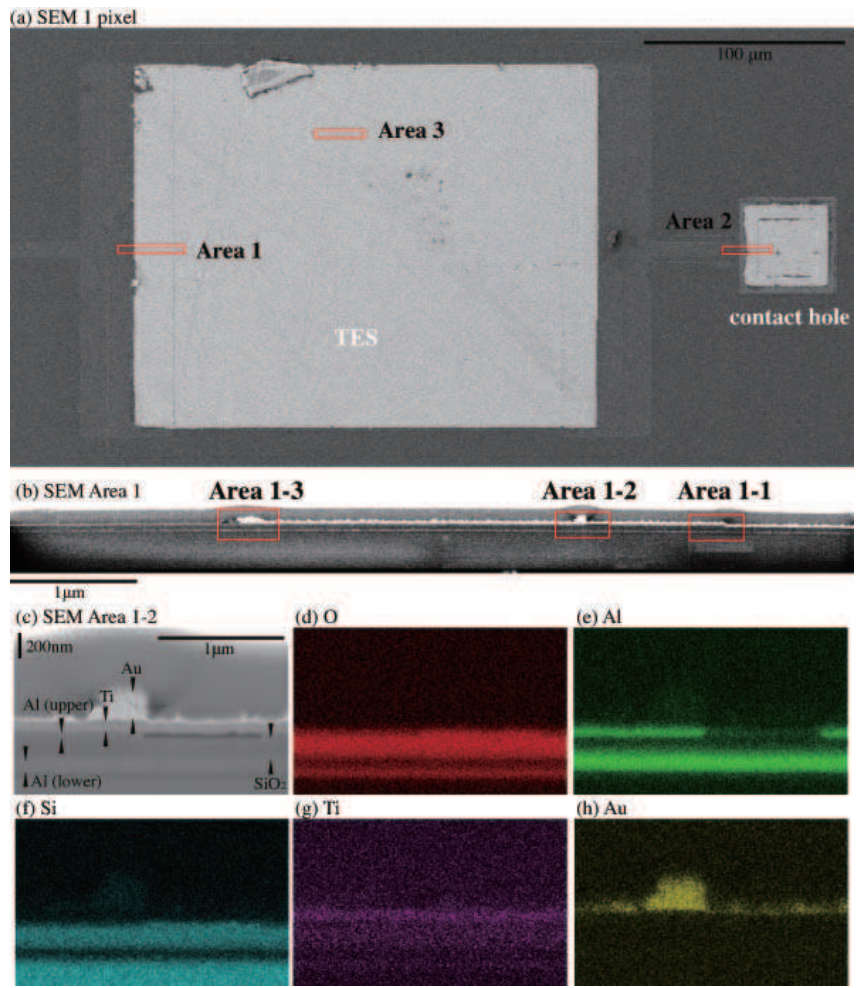}
\end{center}
\caption{SEM and EDS pictures of the cross-section at the contact region between the TES and Al upper wire and the content hole, taken from one of the multi-layer wiring 20$\times$20 TES pixels. 
(a) SEM top-view picture of one pixel. (b) SEM cross-sectional view at Area 1. 
(c) SEM cross-sectional view at Area 1-2. The others are EDS pictures identified as 
(d) Oxygen, (e) Aluminum, (f) Silicon, (g) Titanium,  and (h) Gold.  (color figure online)
}
\label{quad}
\end{figure}
To further study these differences, 
we scrutinized cross-sectional views of the 20$\times$20 TES 
by using Scanning Electron Microscope (SEM) and Focusing Ion Beam (FIB) with Energy Dispersive X-ray Spectrometer (EDS). 
Figure 4a shows a top view of SEM picture taken from one pixel of the 20$\times$20 TES, 
where we marked three portions according to our interest: 
the junction between Al wire and TES (Area 1), 
the contact hole (Area 2), 
and the middle portion of the TES (Area 3). 
In the latter two parts, the TES layers and wires showed good homogeneity and adhesiveness. 

There was, however, found serious issues in Area 1. 
Figure 4b shows the cross-sectional view of the multi-layer wiring TES pixel of Area 1; furthermore, SEM and EDS pictures in Area 1-2 are shown in figure 4c--4h. 
This portion should have maintained Al-upper wire by 100 nm thick and Au (TES) by 250 nm thick,  
though the Al-upper wire seems to have huge void (figure 4e) and the Au layer are far thinner than 250 nm (figure 4h). 
To identify a fabrication process that affected these features, 
we made another multi-layer wiring 20$\times$20 TES pixels, and measured R-T curves for each process step. 
Eventually, we noticed that normal resistance increased from $\sim$ 100 m$\Omega$ to $\sim \Omega$ after TES pattering process. Therefore, it is likely that something during TES patterning might affect the lack of Al and Au. 
One of the key countermeasures is to make the TES layer thinner while the Al layer thicker.
As a next step, we are trying
to fabricate taper-shaped Al upper wires, 
because smooth edges of Al wire make heights of TES layers shorter, while 
the hight of Al higher, 
presumably making the transition temperature lowered and 
the contact between TES with Al wires more smooth and robust, 
resulting in better energy resolution and anticipated superconducting behavior. 

In summary,   
we have successfully obtained X-ray signals from the 4x4 pixel multi-layer readout TES, 
in order to fabricate large array TES used for the future X-ray missions, such as DIOS. 
We found that some of fabrication processes that worked for 4x4 TES were not directly applicable to the 20x20 TESA.Further improvement are still underway, including changing edge shape of Al wires.  

\begin{acknowledgements}

The author would like to thank for many students involved in this work. 
The research presented here has been financed by
the Special Postdoctoral Researchers Program in RIKEN
and JSPS KAKENHI Grant Number 24740129.

\end{acknowledgements}


\end{document}